\documentclass{aa}

\usepackage{graphicx}
\usepackage{placeins}
\usepackage{txfonts}
\usepackage{natbib}
\usepackage{subfigure}
\usepackage{color}
\begin{document} 

   \title{What the Milky Way bulge reveals about the initial metallicity gradients in the disc}
   

   \author{F. Fragkoudi
          \inst{1}
          \and
          P. Di Matteo\inst{1}
           \and
          M. Haywood\inst{1} 
          \and
          S. Khoperskov\inst{1}
          \and
          A. Gomez\inst{1} 
          \and
          M. Schultheis\inst{2}
          \and
          F. Combes\inst{3,4}
          \and
          B. Semelin\inst{3}
          }

   \institute{GEPI, Observatoire de Paris, PSL Research University, CNRS, Place Jules Janssen, 92195,\\
Meudon, France\\
              \email{francesca.fragkoudi@obspm.fr}
         \and
               Laboratoire Lagrange, Universit\'{e} C\^{o}te d'Azur, Observatoire de la C\^{o}te d'Azur, CNRS, Bd de l'Observatoire, 06304 Nice, France
         \and
             Observatoire de Paris, LERMA, CNRS, PSL Univ., UPMC, Sorbonne Univ., F-75014, Paris, France
         \and
             College de France, 11 Place Marcelin Berthelot, 75005, Paris, France
             }

   \date{}


  \abstract
   {We use APOGEE
DR13 data to examine the metallicity trends in the Milky Way (MW) bulge   and we explore their origin by comparing two N-body models of isolated galaxies that develop a bar and a boxy/peanut (b/p) bulge. Both models have been proposed as scenarios for reconciling a disc origin of the MW bulge with a negative vertical metallicity gradient. The first model is a superposition of co-spatial, i.e. overlapping, disc populations with different scale heights, kinematics, and metallicities. In this model the thick, metal-poor, and centrally concentrated disc populations contribute significantly to the stellar mass budget in the inner galaxy. The second model is a single disc with an initial steep radial metallicity gradient; this disc is mapped by the bar into the b/p bulge in such a way that the vertical metallicity gradient of the MW bulge is reproduced, as has been shown already in previous works in the literature. However, as we show here, the latter model does not reproduce the positive longitudinal metallicity gradient of the inner disc, nor the metal-poor innermost regions seen in the data. On the other hand, the model with co-spatial thin and thick disc populations reproduces all the aforementioned trends. We therefore see that it is possible to reconcile a (primarily) disc origin for the MW bulge with the observed trends in metallicity by mapping the inner thin and thick discs of the MW into a b/p. For this scenario to reproduce the observations, the $\alpha$-enhanced, metal-poor, thick disc populations must have a significant mass contribution in the inner regions, as has been suggested for the Milky Way. }

   \keywords{Galaxy: bulge - Galaxy: disc - Galaxy: structure
               }

   \maketitle
   \titlerunning
   \authorrunning

%

\section{Introduction}
\begin{figure*}
\centering
\includegraphics[width=0.95\linewidth]{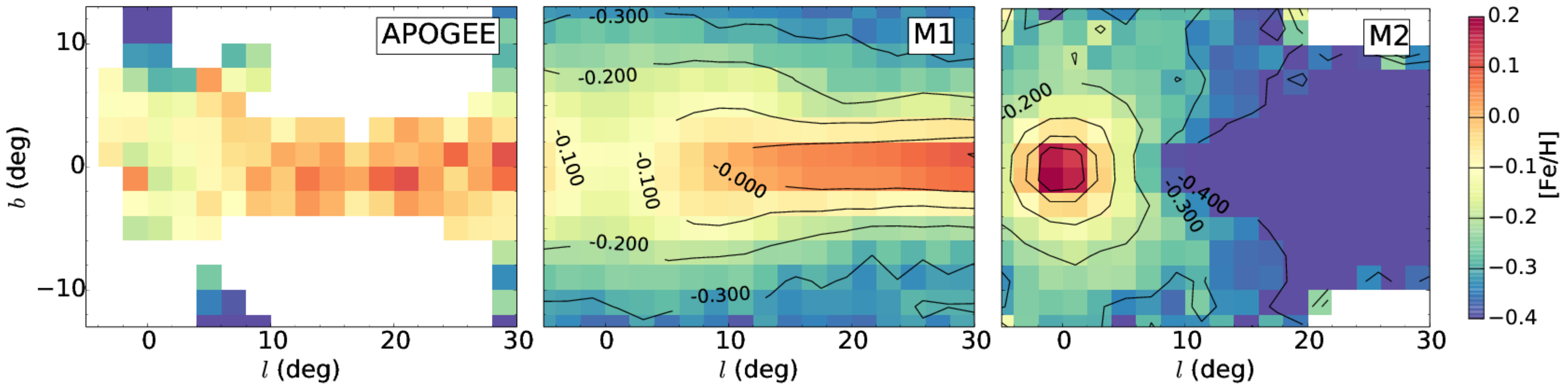}
\caption{Mean metallicity along the line of sight as a function of Galactic longitude $l$ and Galactic latitude $b$ for the APOGEE DR13 data (left), model \texttt{M1} (middle) and model \texttt{M2} (right). For the observed (theoretical) map only stars (particles) with distances between 4 and 12\,kpc from the Sun are selected. Only pixels with more than 10 stars (particles) are shown. Each bin is 2 degrees in $l$ and $b$. The black contour lines correspond to iso-metallicity contours, as labelled on the colour bar.}
\label{fig:meanmetallicity}
\end{figure*}

By examining the morphology, chemistry, and kinematics of the Milky Way (MW) bulge, we  hope to disentangle its formation history (e.g. \citealt{Caluraetal2012, Obrejaetal2013}) and the origin of its various stellar populations, as revealed by its broad metallicity distribution function (MDF; \citealt{McWilliamRich1994,Hilletal2011,Nessetal2013a, RojasArriagadaetal2014,Zoccalietal2017}).

In terms of morphology it is well established that the MW bulge has a characteristic X-shape, as can be seen from images in the near- and mid-infrared \citep{Dweketal1995,NessLang2016} and as is evidenced by the split in the red clump magnitude distribution \citep{McWilliamandZoccali2010}. This is interpreted as due to the presence of a boxy/peanut (b/p) bulge, a structure that forms from vertical heating of stellar bars through resonances and/or the buckling instability \citep{CombesSanders1981,Combesetal1990,Rahaetal1991, Athanassoula2005,MartinezValpuestaetal2006,Quillenetal2014}. 
The MW bulge is therefore thought to have, at least partly, a disc origin, and is made up of material from within the inner disc out to the outer Lindblad resonance (OLR, \citealt{DiMatteoetal2014,Halleetal2015}); values for the present day OLR range between $\sim$8-10\,kpc (see e.g. \citealt{BlandHawthornGerhard2016}).

Regarding the chemistry of the MW bulge, it is now well known that it has a negative vertical metallicity gradient (\citealt{Minnitietal1995,Zoccalietal2008,Johnsonetal2011,Gonzalezetal2013}) due to the changing contribution of various stellar populations above the plane; i.e. the fraction of metal-poor stars ([Fe/H] $\leq$ 0) increases as we move further away from the plane of the Galaxy, while the fraction of metal-rich stars ([Fe/H] $\geq$ 0) decreases (e.g. \citealt{Nessetal2013a,RojasArriagadaetal2014}). 
This vertical metallicity gradient was originally thought to be incompatible with a pure disc scenario for the MW bulge and a dispersion-dominated classical bulge/spheroid was invoked to explain it (e.g. \citealt{Zoccalietal2008}). However, subsequent work has shown that the kinematics of the MW bulge are incompatible with the presence of a \emph{massive} classical bulge, which would be required to explain the vertical metallicity gradient (e.g. \citealt{Shenetal2010, Kunderetal2012,DiMatteoetal2014, Debattistaetal2017}; upper limit between 2-10\% of the total stellar mass). While \cite{Sahaetal2012} showed that classical bulges can be spun up by bars, it is not evident that \emph{massive} classical bulges could acquire the rotation required to hide them kinematically (see e.g. \citealt{Fux1997,DiMatteoetal2014}).

\cite{MartinezValpuestaGerhard2013} showed that it is possible to reconcile the vertical gradient with a disc origin of the bulge when the pre-existing disc has a steep initial radial metallicity gradient (see also \citealt{BekkiTsujimoto2011}); they showed that this steep radial gradient is mapped\footnote{Here and in what follows,  bar "maps" stars into the b/p means that disc stars are redistributed by the bar into the b/p bulge.} by the bar into the b/p in such a way that the bulge recovers a vertical metallicity gradient. This is because while stars from most of the disc participate in the formation of the b/p, stars from larger galactocentric radii preferentially get mapped into the bulge at larger heights above the plane (\citealt{MartinezValpuestaGerhard2013,DiMatteoetal2014}). \cite{DiMatteoetal2015} later showed that while the above scenario can reproduce the global kinematic properties of the MW bulge, this model does not reproduce the kinematics of separate stellar populations, i.e. when the stars are separated according to their metallicity (see \citealt{DiMatteoetal2015} for more details). 

Another recently proposed scenario for reconciling the disc origin of the MW bulge with its vertical metallicity gradient is one in which the metal-poor, $\alpha$-enhanced stars in the bulge are part of the same population as the thick disc stars at the solar vicinity \citep{BekkiTsujimoto2011, Haywoodetal2013,DiMatteoetal2015, DiMatteo2016}; thus the vertical metallicity gradient is present in the disc initially, and is enhanced through vertical heating by the bar. This scenario also reproduces the morphological and kinematic properties of the MW bulge \citep{Athanassoulaetal2017, Debattistaetal2017, Fragkoudietal2017}. In this framework, the $\alpha$-enhanced thick disc would have to be overall massive (as asserted by \citealt{Snaithetal2015,Haywoodetal2015}) and centrally concentrated, i.e. with a short scale length (as shown by \citealt{Bensbyetal2011,Bovyetal2012} and see table \ref{tab:info}) and therefore would be a dominant component of the inner MW. In this context, the bulge is not an exclusively old population, but rather has a spread in ages - as has indeed been suggested by recent studies (e.g. \citealt{Bensbyetal2013,Haywoodetal2016b,Bensbyetal2017}).

In this letter we explore the metallicity trends arising in the two aforementioned ``pure disc'' models and compare these trends with data from the infrared APOGEE survey \citep{Majewskietal2015}, which can probe stellar populations close to the plane of the inner MW. We show that while the model with a single disc and an initial steep radial gradient can reproduce the vertical gradient of the MW bulge, it cannot reproduce the longitudinal gradient in the inner MW nor the average metallicity in the innermost regions. This hints at the lack of an initial steep radial metallicity gradient in the MW disc at the time the bar buckled to form the b/p bulge. On the other hand, we show that the model with co-spatial thin and thick discs, which are mapped into the bulge by the bar, can naturally reproduce all the aforementioned trends; in particular, the low mean metallicity ([Fe/H] $\sim$ -0.1) observed in the inner few degrees of the MW bulge can be reproduced in a pure (thin+thick) disc scenario with no real need for any additional massive component, such as a classical bulge.

\section{Simulations}
\label{sec:5models}

\begin{figure*}
\centering
\includegraphics[width=0.75\linewidth]{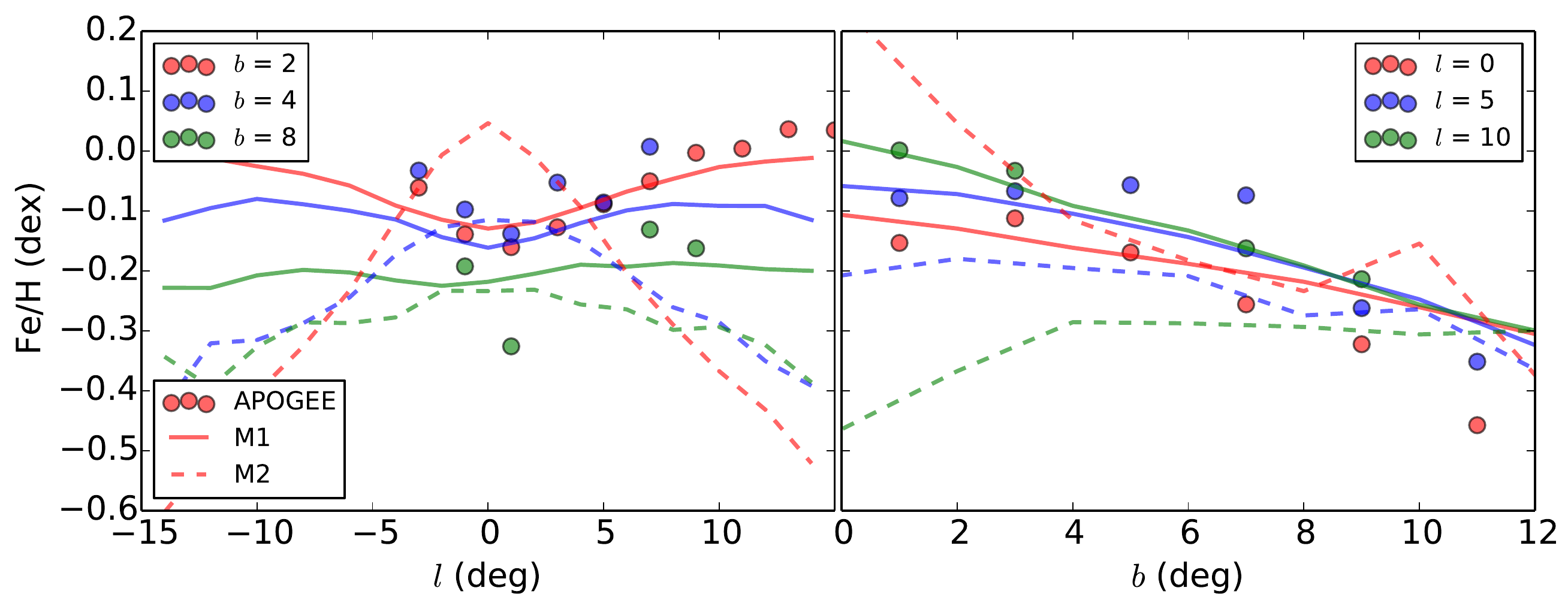}
\caption{\emph{Left:} Mean metallicity as a function of longitude for various cuts in latitude as indicated in the legend. \emph{Right:} Mean metallicity as a function of latitude for various cuts in longitude as indicated in the legend. The solid line corresponds to Model \texttt{M1} while the dashed line corresponds to Model \texttt{M2}. The circles correspond to data from APOGEE. For the observations (models) only stars (particles) with distances
between 4 and 12\,kpc from the Sun are considered.} 
\label{fig:slits}
\end{figure*}
\subsection{\emph{\texttt{M1}:} Co-spatial discs}

In this model\footnote{The main properties of the two models are summarised here. For a more detailed description see Appendix \ref{sec:appendixIC}.} we discretise the vertically continuous stellar populations seen in the MW inner disc (see \citealt{Bovyetal2012}, Figure 5) into three co-spatial i.e. overlapping, discs: \emph{D1}) a thin, kinematically cold, metal-rich disc ([Fe/H] > 0), which is associated with the metal-rich thin disc seen in the solar vicinity; \emph{D2}) an intermediate disc with intermediate kinematics and metallicities (0 > [Fe/H] > -0.5), which is associated with the young thick disc; and \emph{D3}) a kinematically hot and metal-poor disc (-0.5 > [Fe/H] > -1) with scale height and kinematic properties corresponding to the old thick disc seen at the solar vicinity (nomenclature as in \citealt{Haywoodetal2013}). 

Disc particles are assigned a metallicity by drawing randomly from a normal distribution, where each disc has a mean metallicity and dispersion (see table \ref{tab:info}) assigned such that we can reproduce the MDF of the inner MW (Haywood et al., submitted). Discs \emph{D2} and \emph{D3} are assigned a flat radial metallicity gradient, since we assume that the interstellar medium (ISM) was well mixed at the time the young and old thick disc formed ($z$\,>\,1) owing to a high star formation rate (>\,10 $M_{\odot}$/yr), meaning that  the galaxy was likely in a bursty and turbulent state (\citealt{Lehnertetal2014,Wuytsetal2016, Maetal2017}, but see also \citealt{Pilkingtonetal2012}). The thin disc (\emph{D1}) is associated with the final more quiescent phase of the MW in the last $\sim$7-8 Gyr (see \citealt{Snaithetal2014,Haywoodetal2016}). We are only interested in the gradient of the \emph{inner} thin disc, since the outer disc does not participate in the b/p bulge. While it is possible that the inner thin disc could have a negative gradient, we investigate the simplest case first, i.e. a flat gradient. 
The combined mass of \emph{D2} and \emph{D3} (young and old thick) discs is $\sim$50\% of the total stellar mass in the model and the other $\sim$50\% is in the thin disc component, which is in agreement with the mass growth of the MW disc as estimated by \cite{Snaithetal2014,Snaithetal2015}. 

The fact that \emph{D2} and \emph{D3} have a larger scale height than the thin disc leads to a global negative vertical metallicity gradient at the start of the simulation (see left column Figure \ref{fig:seq}). The model represents the inner MW, i.e. up to the OLR (see e.g. \citealt{BlandHawthornGerhard2016}). The model is evolved over secular timescales, i.e. 7\,Gyr until a bar and b/p form, and in what follows, we analyse a snapshot from the end of the simulation, which is re-scaled so that the bar has a length of $\sim$4.5\,kpc that is similar to the length of the MW bar (e.g. \citealt{BlandHawthornGerhard2016}).

\subsection{\emph{\texttt{M2}:} Radial metallicity gradient}

In this model (\texttt{M2}) we employ a single stellar disc that has an initial steep radial metallicity gradient. The particles are assigned a metallicity as a function of their initial radius in the disc according to [Fe/H] = [Fe/H]$_0$ + grad $\times$ $r$, where grad = -0.4\,dex/kpc and [Fe/H]$_0$ = 0.6\,dex to emulate the models explored in \cite{MartinezValpuestaGerhard2013} and \cite{DiMatteoetal2015}. The model is evolved over secular timescales for 9\,Gyr and in what follows we analyse the final snapshot, which is re-scaled so that the bar has a length of $\sim$4.5\,kpc -- similar to the MW bar length.

\section{Results} 
\label{sec:metals}

In Figure \ref{fig:meanmetallicity} we explore the mean metallicity [Fe/H] along the line of sight in Galactic longitude $l$, and Galactic latitude $b$, for the two models described above. We select particles that are in the bar/bulge region with distances between 4 and 12\,kpc from the Sun, which is placed at 8\,kpc from the Galactic centre and the bar has an angle of 30 degrees to the galactocentric line of sight. 
We compare these to a mean metallicity map of the inner MW constructed with data from APOGEE DR13 \citep{Majewskietal2015,DR132016}, where stars are selected to have distances between 4 and 12\,kpc from the Sun. The distances are taken from \cite{Wangetal2016} (see also their section 3 for more details on how stars in this sample are selected), and we apply an additional cut on surface gravity, log\,$g$ > 0.5, to remove biases from the most distant metal-poor stars in the bulge (see section 5.1 in \citealt{Nessetal2016}). Only bins with more than 10 stars are shown (bins are 2 $\times$ 2 deg) and the total number of stars used to construct this map is $\sim$7300.

We see that model \texttt{M1} (middle panel Figure \ref{fig:meanmetallicity}) is able to reproduce the vertical metallicity gradient, seen also in the APOGEE data (left panel Figure \ref{fig:meanmetallicity}), due to the changing contribution of the different populations with latitude (see the right column of Figure \ref{fig:frac}, where we show the fractional contribution of the three co-spatial populations in $l,b$ for this snapshot). 
Model \texttt{M2} (right panel Figure \ref{fig:meanmetallicity}) also reproduces the vertical metallicity gradient because stars that are further out in the disc are mapped at larger distances above the plane by the bar in the b/p bulge (see \citealt{MartinezValpuestaGerhard2013} and \citealt{DiMatteoetal2014}). 
However, the two models give very different predictions for the longitudinal metallicity gradient. We see that the inner disc of \texttt{M1}  (30  > $l$ > 10) is metal-rich close to the plane. This is due to the thin, metal-rich thin disc, which is concentrated in the plane of the galaxy (it contributes $\sim$60\% of the surface density for $b$<|5|; see right column of Figure \ref{fig:frac}). 
On the other hand, in Figure \ref{fig:meanmetallicity}, we see that \texttt{M2} predicts a metal-poor inner disc at longitudes $l$ > 10 degrees. These are diametrically opposite predictions in the sense that \texttt{M1} predicts a positive longitudinal gradient, while \texttt{M2} predicts a negative longitudinal gradient, and we see from the APOGEE data that \texttt{M1} can reproduce the data much better than \texttt{M2}. 

Furthermore, the models also give different predictions for the mean metallicity in the inner regions; \texttt{M1} predicts relatively low metallicity inside |$l,b$| < 5 deg, of the order of -0.1 dex, while \texttt{M2} predicts that metallicity is highest in the centre, of the order of 0.1dex. We see that \texttt{M1} is compatible with the APOGEE data, which points to low metallicities of on average $\sim$\,-0.1\,dex (and see also \citealt{Zoccalietal2017}).

The relation between the models and data can be further explored by examining Figure \ref{fig:slits}. In the left panel we show metallicity as a function of longitude for various cuts in $b$ as indicated in the inset. We see that \texttt{M1} (solid lines) predicts increasing metallicity towards larger longitudes with a dip in the centre (at $l$=0), whereas \texttt{M2} (dashed lines) predicts decreasing metallicity with a peak in the centre. For the APOGEE data (circles) the error on the mean for the data is smaller than the points and therefore is not shown. We see that there is a very good agreement between the trends for model \texttt{M1} and the data, especially close to the plane. i.e. for $b$=2, while model \texttt{M2} is essentially excluded. The data seem to show an inversion of metallicities compared to the models, around $l$=0, for $b$ = 2 and 4; i.e. $b$=2 seems to be more metal-poor than $b$=4, contrary to the prediction of both the models. Since both models are pure discs, this could possibly hint at the existence of a small, concentrated classical bulge or a contribution from the stellar halo, which make the very innermost region of the MW more metal poor. In the right panel of Figure \ref{fig:slits} we show the metallicity as a function of $b$ for various cuts in longitude, where the negative vertical gradient is clearly seen in the data. 
We note that the metallicity in model \texttt{M1} does not decrease as rapidly as in the data owing to a larger fraction of metal-rich stars at high latitudes. This occurs because we do not have star formation in the model and thus all cold, metal-rich populations are present before the bar buckling episode. Notwithstanding, the main trends in the data are well reproduced by the model.



\section{Discussion and conclusions}
\label{sec:discussion}

We examine metallicity trends in the Milky Way bulge using APOGEE DR13, and compare the data to two pure disc models, both able to reproduce the MW bulge's vertical metallicity gradient. The first model (\texttt{M1}) has thin and thick disc stellar populations, which are represented by a thin, kinematically cold metal-rich disc ([Fe/H] > 0); an intermediate disc in terms of thickness, kinematics, and metallicity (-0.5 < [Fe/H] < 0); and a thicker, kinematically hotter metal-poor disc (-1 < [Fe/H] < -0.5), where the intermediate and thick discs make up approximately 50\% of the stellar mass and are centrally concentrated. The second model (\texttt{M2}) has a single disc with an initial steep radial metallicity gradient (-0.4\,dex/kpc); Both models are evolved in isolation and develop a bar that buckles to form a boxy/peanut bulge.

While the bar in model \texttt{M2} maps the initial steep radial gradient into a negative vertical metallicity gradient in the b/p bulge -- thus producing a negative vertical gradient (as shown in \citealt{BekkiTsujimoto2011,MartinezValpuestaGerhard2013} and \citealt{DiMatteoetal2014}) -- it however fails to reproduce the positive longitudinal metallicity gradient close to the plane and the mean metallicity in the innermost regions of the MW bulge (inside |$l,b$| < 5 deg). These regions close to the plane can now be observed thanks to the infrared APOGEE survey. On the other hand, we see that \texttt{M1} is able to reproduce all the aforementioned trends in the data. 

The inability of model \texttt{M2} to reproduce the trends in the MW bulge hints at the lack of a steep radial metallicity gradient in the inner MW at the time the bar buckled to form the b/p bulge. This is in line with recent observational (e.g. \citealt{Wuytsetal2016}) and theoretical studies that show that a large fraction of galaxies at high redshifts have flat gas-phase metallicity gradients due to strong feedback and disc-disturbing processes such as rapid gas infall (e.g. \citealt{Gibsonetal2013, Maetal2017}). 
While we do not exclude that the inner MW could have had an initial shallow negative radial gradient, according to our results, it had to be flatter than that of \texttt{M2}. On the other hand, if the initial radial gradient is less steep than in model \texttt{M2}, the vertical gradient produced in the b/p bulge is not as steep as required by the data. Therefore, there must be another scenario that is able to produce the fairly steep negative vertical gradient observed in the MW bulge, while also satisfying the rest of its chemical, morphological, and kinematic trends. 

As mentioned, model \texttt{M1} is able to naturally reproduce both the vertical and longitudinal metallicity gradients seen in the APOGEE data, as well as the metal-poor inner region (|$l,b$| < 5 deg) of the MW bulge. Our findings suggest that the stellar populations that make up the inner MW arose from an ISM that was well mixed and turbulent and whose radial metallicity gradients were mostly flat \citep{Haywoodetal2013,Nideveretal2014,FengKrumholz2014,DiMatteoetal2015,Haywoodetal2015,Wuytsetal2016,DiMatteo2016}, with stars first forming in a geometrically thick layer and then in thinner layers in an upside-down fashion (e.g. \citealt{Birdetal2013}). In such a model, where the MW bulge forms via the bar mapping the thin and thick inner disc populations into a b/p bulge, it is not necessary to add any other massive components (such as a classical bulge/spheroid) to the model to reproduce the trends in metallicity, nor is an initial radial metallicity gradient in the disc needed.

\begin{acknowledgements}
This work has been supported by the ANR (Agence Nationale de la Recherche) through the MOD4Gaia project (ANR-15-CE31-0007, P.I.: P. Di Matteo). FF is supported by a postdoctoral grant from the Centre National d'Etudes Spatiales (CNES). This work was granted access to the HPC resources of CINES under the allocation 2016-040507 made by GENCI. 
Funding for SDSS-III has been provided by the Alfred P. Sloan Foundation, the Participating Institutions, the National Science Foundation, and the U.S. Department of Energy Office of Science. The SDSS-III web site is http://www.sdss3.org/.
\end{acknowledgements}

\bibliographystyle{aa}
\bibliography{References}%

\begin{appendix} 
\section{Initial conditions \& codes}
\label{sec:appendixIC}
The initial conditions of the two models explored in this study are obtained using the iterative method of \cite{Rodionovetal2009}. The algorithm constructs equilibrium phase models for stellar systems with a constrained evolution so that the equilibrium solution has a number of desired parameters. In our case, these parameters are the density distribution of the discs (see table \ref{tab:info}), which are modelled as Miyamoto-Nagai profiles, and the density distribution of the dark matter halo, which is modelled as a Plummer sphere \citep{BT2008}.
\paragraph{Simulation \texttt{M2}} was run with the Tree-SPH code of \cite{SemelinCombes2002}, in which gravitational forces are calculated using a hierarchical tree method \citep{BarnesHut1986}. A Plummer potential is used to soften gravity at scales smaller than $\epsilon$ = 150\,pc. The equations of motion are integrated using a leapfrog algorithm with a fixed time step of $\Delta t$ = 0.25\,Myr. The live dark matter halo has 5$\times 10^5$ particles and a total mass of 1.61 $\times$ $10^{11}$ with a Plummer radius of 10\,kpc, and the disc has 1$\times 10^6$ particles. For a full description of the code, see \cite{SemelinCombes2002}.

\paragraph{Simulation \texttt{M1}} was run with a recently developed parallel MPI Tree-code, which takes into account the adaptive spatial decomposition of particle space between nodes. The multi-node Tree-code is based on the 256-bit AVX instructions, which significantly speed up the floating point vector operations and sorting algorithms (Khoperskov et al. in prep). The softening for model \texttt{M1} is  $\epsilon$ = 50\,pc. In total, the disc components of \texttt{M1} have 10$\times 10^6$ particles and the live dark matter halo is assigned 5$\times 10^6$ particles with a total mass of 3.68 $\times$ $10^{11}$ with a Plummer radius of 21\,kpc. 

\begin{table*}
\centering
\label{tab:info}
\begin{tabular}{ l  r | c | c | c | c | c | c  } 
\texttt{M1} & & $r_D$ (kpc) & $h_z$ (kpc)  & $M$ ($M_{\odot}$) & $n_p$  & [Fe/H] (dex) & $\sigma_{[Fe/H]}$ (dex) \\ \hline
&\emph{Thin (\emph{D1})} & 4.8 & 0.15 & 4.21 $\times$ $10^{10}$ & 5000000  & 0.3 & 0.15 \\ \hline
&\emph{Intermediate (\emph{D2})} & 2 & 0.3 & 2.57 $\times$ $10^{10}$ & 3000000 & -0.3 & 0.2 \\ \hline
&\emph{Thick (\emph{D3})} & 2 & 0.6 & 1.86 $\times$ $10^{10}$ & 2000000  & -0.65 & 0.2 \\ \hline
\texttt{M2} &  \\ \hline
& \emph{Thin} & 4.7 & 0.3 & 10 $\times$ $10^{10}$ & 1000000  & 0.6 - 0.4 $\times$ $r$ & -   \\
\end{tabular}
\vspace{0.1cm}
\caption{Properties of the simulations used in this study. From left to right: The characteristic radius of the population, the characteristic height of the population, mass of the component, number of particles in component, the mean metallicity of the component, and the dispersion in metallicity.}
\end{table*}

\section{Evolution in time}
\label{sec:appendixB}

\begin{figure*}
\centering
\includegraphics[width=0.8\linewidth]{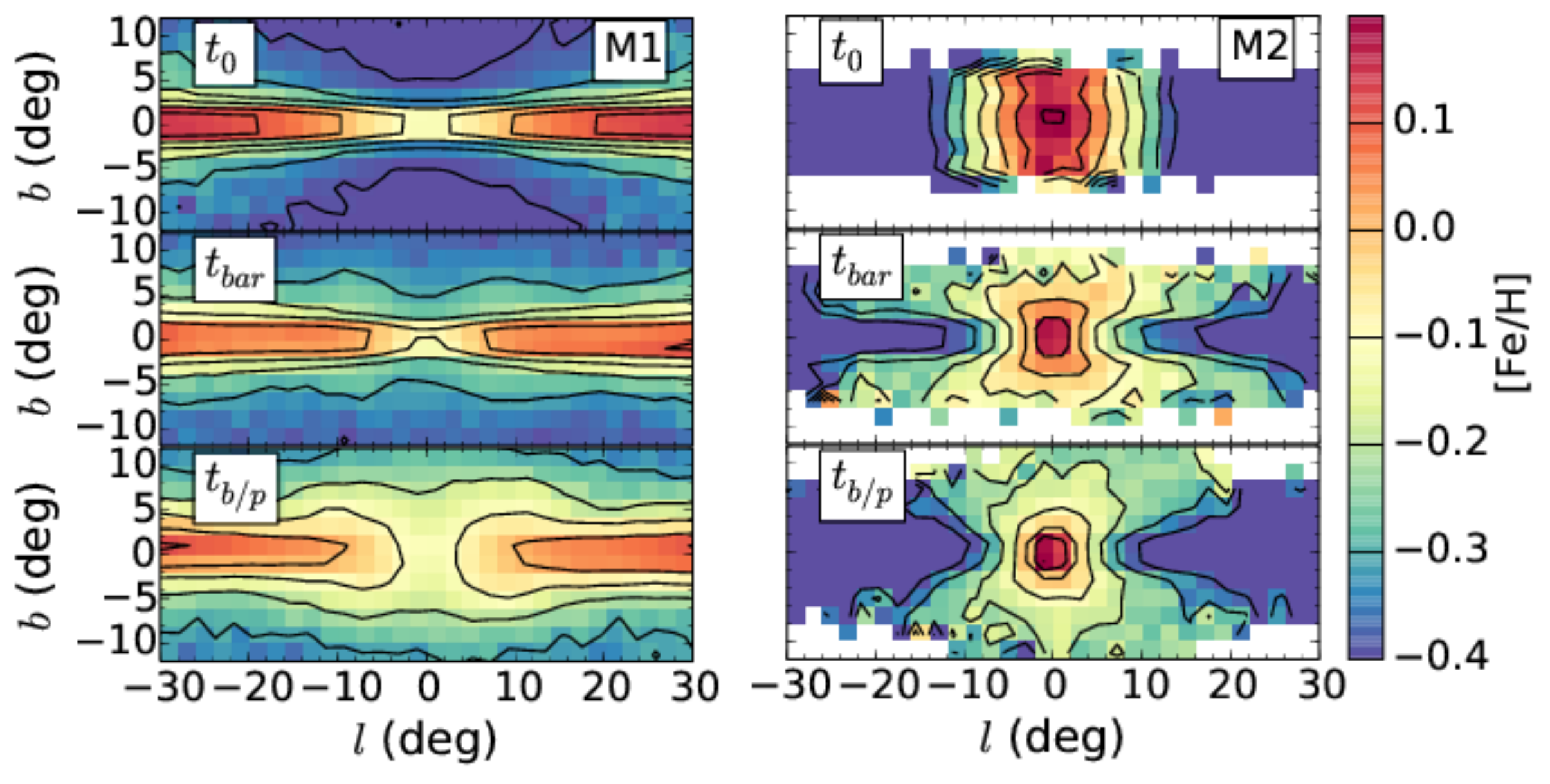}
\caption{Metallicity maps in $l,b$ for model \texttt{M1} (left) and model \texttt{M2} (right) at the start of the simulation $t_0$ (top panels), after bar formation $t_{bar}$=3\,Gyr (middle panel) and after the bar buckles and the b/p forms $t_{b/p}$=5.5\,Gyr (bottom panel). }
\label{fig:seq}
\end{figure*}

\begin{figure*}
\centering
\subfigure[$t_0$]{
        \includegraphics[width=0.4\linewidth]{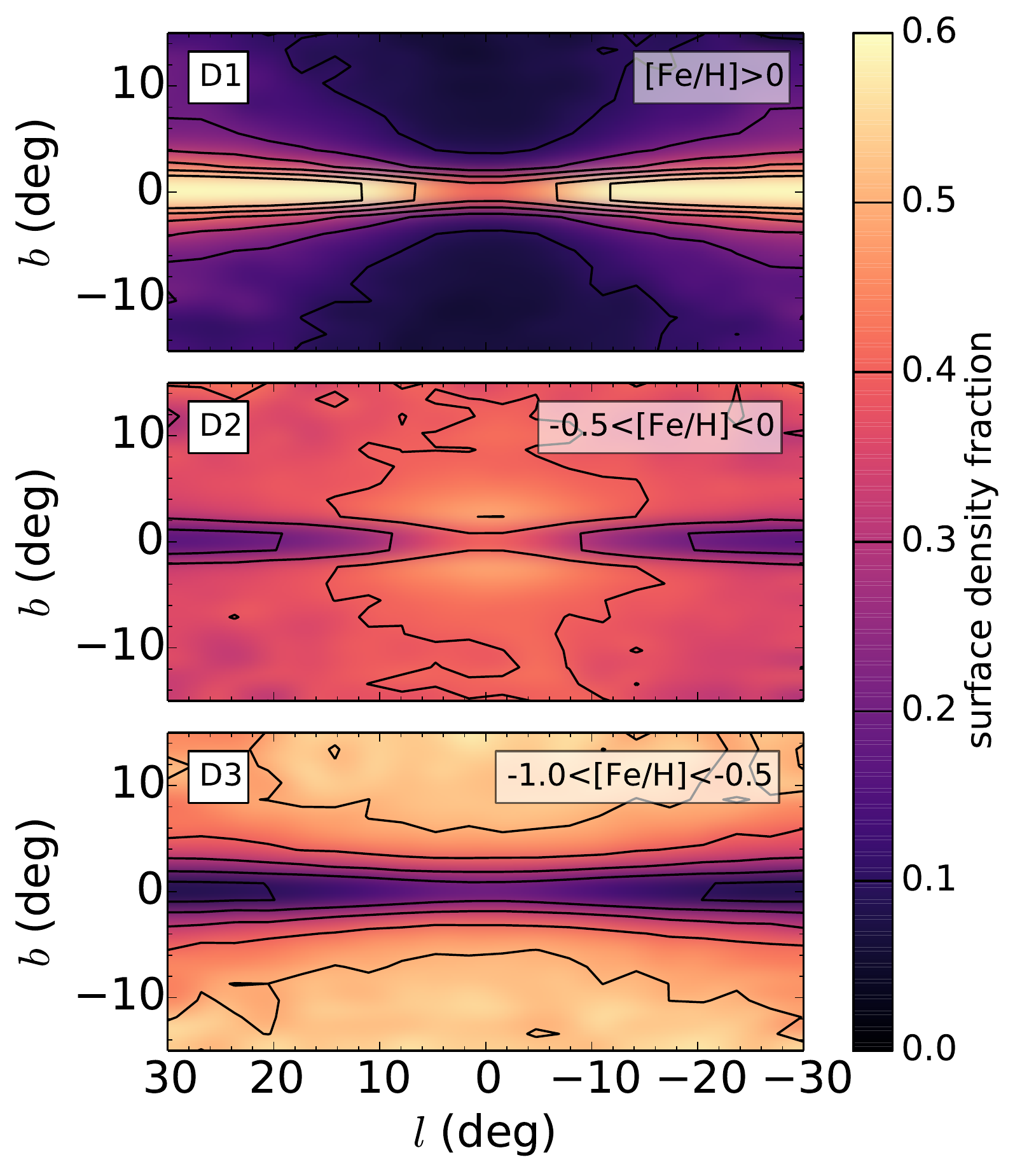}
        \label{fig:1}}
\quad
\subfigure[$t$ = 7\,Gyr]{
        \includegraphics[width=0.4\linewidth]{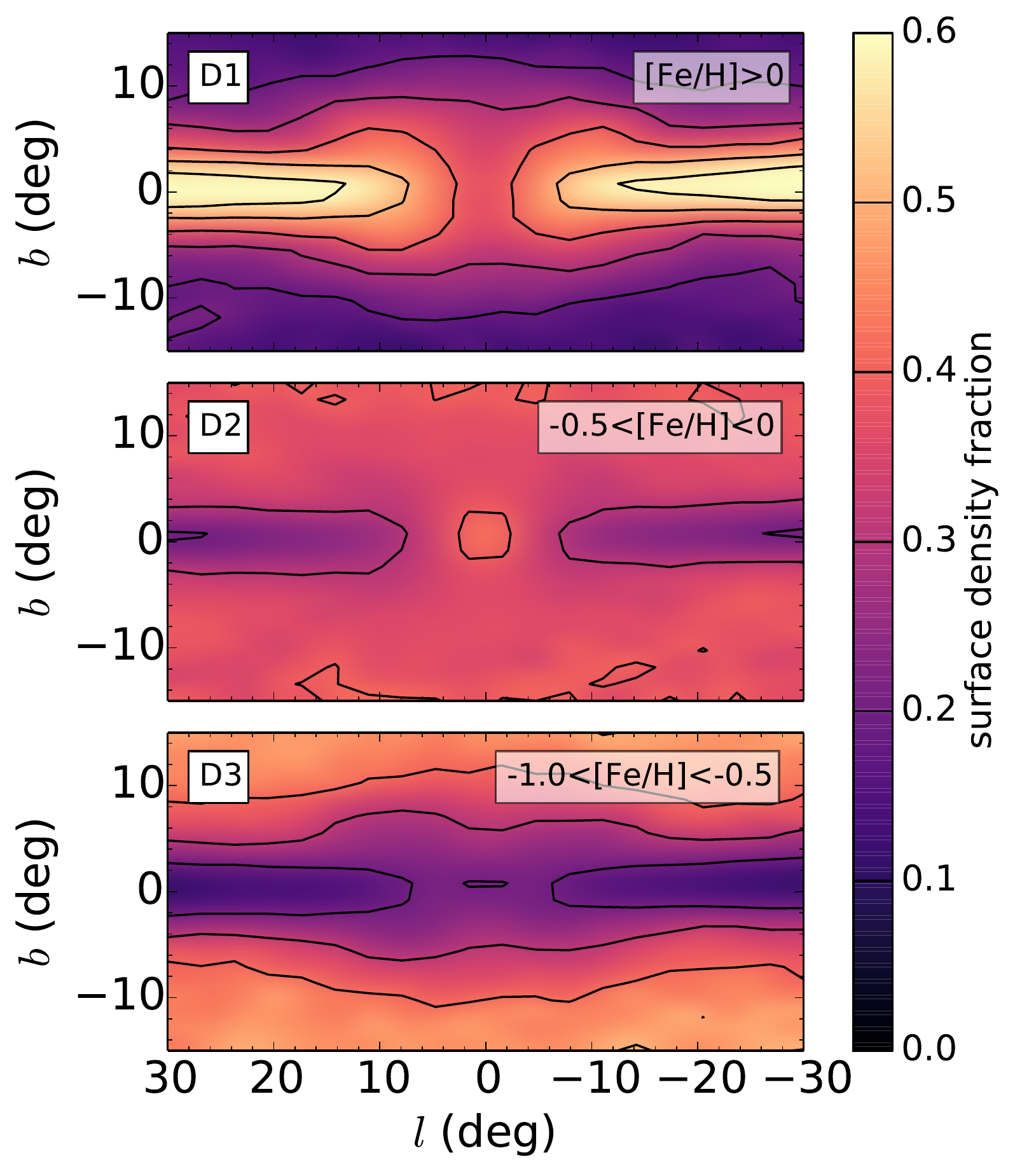}
        \label{fig:2}}
\quad
\caption{For model \texttt{M1}: Fractional contribution of each of the co-spatial discs along the line of sight in $l,b$ for a snapshot at $t$=0 (left) and at 7\,Gyr (right, i.e. the same snapshot as the one shown in Figure \ref{fig:meanmetallicity}), where we chose only particles with distances between 4 and 12\,kpc from the Sun. At late times (right panel) the cold population, \emph{D1}, puffs up and has a distinctive peanut shape.}
\label{fig:frac}
\end{figure*}

In Figure \ref{fig:seq} we show metallicity maps in $l,b$ for the two models for three different snapshots: at the start of the simulation (top panels), after the bar forms (middle panels), and after the b/p bulge forms (bottom panels). We see that for model \texttt{M1} the vertical gradient is present from the beginning of the simulations because of the vertically structured nature of the disc. This vertical negative gradient changes in steepness owing to the formation of the b/p bulge, which increases the heights to which the metal-rich populations reach. In model \texttt{M2,} on the other hand, there is no vertical gradient to begin with and this is induced by the formation of the bar and b/p bulge.

This can be further understood by examining Figure \ref{fig:frac}, where we show the fractional contribution (in surface density) of the three co-spatial discs of model \texttt{M1} before (left panel Figure \ref{fig:frac}) and after (right panel Figure \ref{fig:frac}) the b/p bulge forms. We see that the vertical separation in populations is present from initial times, but that the coldest population (\emph{D1}) is puffed up after the formation of the b/p bulge, as the stars from this population are trapped more strongly in the bar-b/p instability and thus form a more prominent X-shape.

We see that the population \emph{D1}, the thin disc population, is confined close to the plane where it dominates, whereas as we move further away from the plane, the thick disc (\emph{D3}) contribution becomes more and more important especially above latitudes of |$b$|$\sim$10. The contribution from the intermediate disc (\emph{D2}) is approximately constant at all the latitudes explored.

\end{appendix}

\end{document}